%% file: LSSTxEuclid.tex
\documentclass[a4paper,fleqn,usenatbib]{mnras}

\usepackage{newtxtext,newtxmath}

\usepackage[T1]{fontenc}
\usepackage{ae,aecompl}

\usepackage{lipsum}

\usepackage{graphicx}	
\usepackage{amsmath}	
\usepackage{amssymb}	
\usepackage{float}
\usepackage{color}

\definecolor{purple}{RGB}{76, 0,153}

\title[Shape measurement synergies between LSST and Euclid]{Galaxy shape measurement synergies between LSST and Euclid}

\author[R. L. Schuhmann et al.]{
Robert L. Schuhmann$^{1,2}$\thanks{E-mail: roberts@roe.ac.uk (RLS)},
Catherine Heymans$^{1}$,
Joe Zuntz$^{1}$
\\
$^{1}$Institute for Astronomy, University of Edinburgh, Blackford Hill, Edinburgh EH9 3HJ, UK\\
$^{2}$School of Physics and Astronomy, University of Manchester, Manchester M13 9PL, UK\\
}

\date{Accepted XXX. Received YYY; in original form ZZZ}

\pubyear{2019}

\begin{document}
\label{firstpage}
\pagerange{\pageref{firstpage}--\pageref{lastpage}}
\maketitle

\begin{abstract}

\input{abstract_ch.tex}

\end{abstract}




\section{Introduction}
\input{introduction.tex}

\section{Methods}
\label{sec:methods}
\begin{figure}
\centering
 \includegraphics[width=\linewidth,trim=0mm 10mm 0mm 0mm]{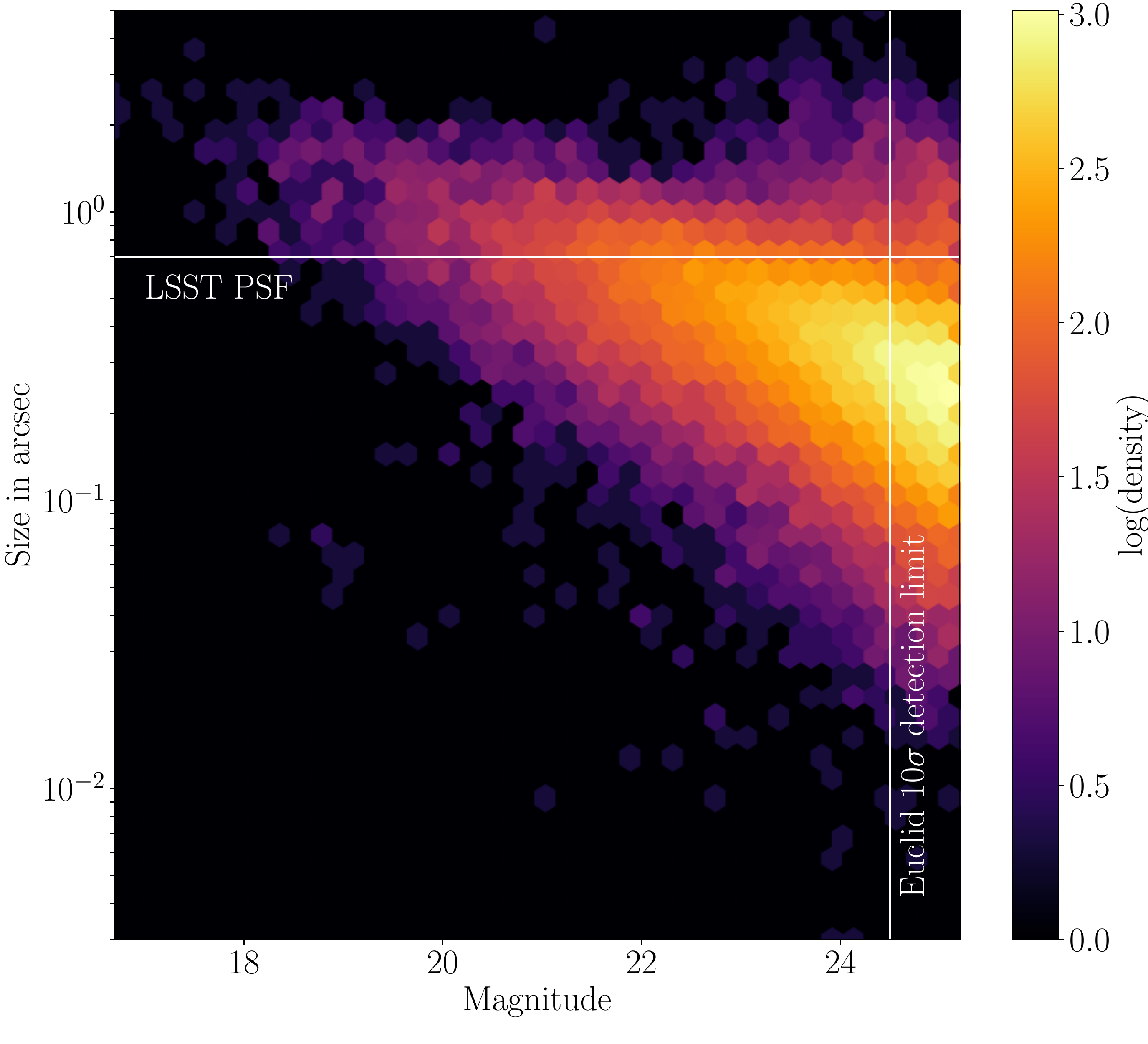}
\caption{2D histogram of measured galaxy sizes and magnitudes (F814W) from the COSMOS survey \citep{Leauthaud/etal:2007}. The horizontal line indicates the typical LSST PSF width of 0.7 arcsec \citep{Chang/etal:2013}.  The vertical line indicates the expected Euclid 10$\sigma$ detection limit of $i \sim 24.5$ \citep{Cropper/etal:2016}.}
\label{fig:histo}
\end{figure}

\subsection{Galaxy Simulations}
\label{sec:simulations}
To simulate LSST-like and Euclid-like galaxy images we use the popular {\sc GalSim}\footnote{See \url{https://github.com/GalSim-developers/GalSim}} image simulation package \citep{Rowe/etal:2015}. This public software allows us to simulate realistic images of galaxies with various models for the point-spread function (PSF) and pixel noise.  As input truth for our simulations, we model the galaxy light profile with a single-component S\'ersic model \citep{sersic:1963}.   The joint S\'ersic index-size-magnitude distribution is then given by deep high resolution observations from the Hubble Space Telescope COSMOS survey \citep[for details of the survey and fitting process see][]{Leauthaud/etal:2007, Lackner/Gunn:2012, Mandelbaum/etal:2014}.   We adopt the random (fair) subsample of the COSMOS galaxies that was used in the GREAT3 galaxy shape measurement challenge, down to a limiting magnitude of F814W<25.2 \citep{Mandelbaum/etal:2014}.  To ensure a realistic size-magnitude distribution for our chosen single-S\'ersic galaxy profile model, we discard 29\% of the objects for which the single-component model provides a significantly worse fit than a two-component bulge+disk profile.  Figure~\ref{fig:histo} shows a joint histogram of magnitudes and galaxy sizes of the resulting 58,074 input objects for our simulations\footnote{We note that the size-magnitude distribution shown in Figure~\ref{fig:histo} is not significantly altered if the best-fitting two-component model galaxies are also included in the galaxy sample.}, where throughout this work, we define {\it galaxy size} to be the half-light radius measured along the major axis.  The distribution of the galaxy population can be compared to the typical resolution for the LSST images (horizontal line) and the expected depth of the Euclid imaging (vertical line).

In this analysis we wish to compare high-resolution limited-depth imaging (Euclid-like) with low-resolution high-depth imaging (LSST-like).   We therefore make the following simplifying assumptions.  We approximate the PSF in both cases to be circular and Gaussian with a width given by the Euclid diffraction limit (0.06 arcsec) or the typical LSST seeing (0.7 arcsec).   The model for the pixel noise consists of  Poisson noise corresponding to the number of electrons in each pixel, including the sky background.  Our chosen values to model the noise are summarized in Table~\ref{tab:galsimpars} (\citealt{Ivezic/etal:2008}, \citealt{Jones:2017}; B. Gillis and D. Kirkby, private communications), which recover the expected Euclid depths as well as the LSST $i$-band signal-to-noise ratio\footnote{Using the exposure time calculator available at \url{https://github.com/jmeyers314/LSST_ETC} \citep{Meyers:2014}}. 
Our selected values for PSF width, exposure time, and CCD noise parameters therefore reflect the underlying guideline to this work: to model the limiting factors of ground-based and space-based imaging, i.e. seeing and pixel noise respectively, and to investigate how their corresponding complementary strengths can compensate for these limitations synergistically.   

To approximate the co-adding process, we simulate a single noisy exposure, multiplying the nominal shutter time for one exposure in each survey (30 sec for LSST; 565 sec for Euclid) with a representative number of the exposures that will be used for one co-added image -- 3 for Euclid and 540 for LSST \citep{Ivezic/etal:2008,LSSTScienceBook/2009,Cropper/etal:2016}. The sky background value is adapted accordingly. We choose the number of LSST exposures to be three times 180, which is the nominal number of $i$-band exposures per pointing -- this is to mimic multi-band coadding. These assumptions are idealisations; a realistic co-adding procedure will not yield results of the same quality. However, for the purposes of our exploratory analysis we deem this prescription sufficient.   As a last step, both noisy exposures are rescaled to be in units of flux per second per square arcsecond, so that the joint pixel measurement is able to use one common parameter for the total flux. 

\begin{table}
\centering
\caption{Simulation Parameters for the LSST-like and Euclid-like images, as used within {\sc GalSim}.}
\begin{tabular}{r|cc}
\hline
parameter & LSST-like & Euclid-like \\
\hline
pixel size (arcsec):& 0.2 & 0.1 \\
exposure size (pixels):& 32$\times$32 & 64$\times$64\\
PSF width (arcsec):& 0.7 & 0.06 \\
effective mirror diameter (m):& 6.423 & 1.13 \\
total exposure time (sec): & 30$\times$540 & 565$\times$3 \\
total sky background (e$^-$/pixel): & 975.7$\times$540 & 114$\times$3\\
\end{tabular}
\label{tab:galsimpars}
\end{table}

\subsection{Shape Measurement}
\label{sec:ngmix}
For each pair of space and ground-based galaxy images, generated from the same COSMOS object, we perform four distinct ellipticity measurements: one LSST-like, one Euclid-like, one by combining these two measurements via a weighted sum (hereafter: ``catalogue combination'', or briefly ``CatComb''), one by fitting one galaxy profile to both images simultaneously (hereafter: ``joint-pixel analysis'', or ``JointPix''). 

We use the shape fitting software {\sc ngmix}\footnote{See \url{https://github.com/esheldon/ngmix}} \citep{Sheldon:2014}, which follows a model-fitting approach. Ideally, we would be using the same shape measurement methods as LSST and Euclid, but their image processing pipelines are still under development. Both are likely to include one or several shape measurements based on model fitting, and {\sc ngmix} has been tested and used extensively within the Dark Energy Survey\footnote{DES: \url{https://www.darkenergysurvey.org/}} (DES -- see \citealt{Jarvis/etal:2016,Zuntz/etal:2017}). Thus we consider our choice of shape measurement software to be a reasonable placeholder for the final pipeline design decisions that LSST and Euclid will make.

{\sc Ngmix} approximates galaxy and PSF profiles as mixtures of concentric Gaussian densities. This model is quick to evaluate as convolutions between the two can be computed analytically \citep{Hogg/Lang:2012}.  In each of the three independent fits (LSST, Euclid, and JointPix), we assume perfect knowledge of the PSFs, which are given by circular Gaussians with the widths set to the values we use to generate the images (see Table~\ref{tab:galsimpars}).  This is highly idealised but captures the essential features of the problem that we are exploring here.

The {\sc ngmix} galaxy model is an exponential profile, described by six parameters $(c_1, c_2, \epsilon_1, \epsilon_2, T, F)$. The first two are the coordinates of the centroid. Further, $T$ is the second moment of the intensity distribution in square arcseconds and $F$ is the total flux of the model divided by the angular area of one pixel, measured in photoelectrons per square arcsecond.
The ellipticity $\epsilon = \epsilon_1 + i \epsilon_2$ quantifies the galaxy shape: if $a$ and $b$ are the major and minor elliptical axes of elliptical isophotes, then the magnitude of the ellipticty is 
\begin{equation}
|\epsilon|=\frac{a-b}{a+b}\leq 1,
\end{equation}
with a phase angle that is twice the orientation angle of the major half axis. 

To demonstrate the synergy benefit for shear calibration, we add a small amount of cosmic shear to each galaxy prior to simulating the galaxy image and measuring the ellipticity. For each galaxy we draw two random numbers $g_1$ and $g_2$ from a normal distribution with mean zero and spread 0.02, and transform the original ellipticity $\epsilon^\text{true}$ via the standard formula
\begin{equation}
	\epsilon^\text{lensed}=\frac{\epsilon^\text{true}+g}{1+g^*\epsilon^\text{true}},
\end{equation}
where we have introduced the complex reduced shear $g=g_1+ig_2$ \citep{SeitzSchneider:1997}. We save both the intrinsic ellipticity $\epsilon^\text{true}$ and the reduced shear $g$ for each object.

Sampling of the 6-dimensional parameter space proceeds in four steps -- this fitting procedure is very similar to the process implemented in \cite{Sheldon:2014}, and is a simplified version of the method used in  \cite{Jarvis/etal:2016}:
\begin{itemize}
\item To reduce runtime, we initialise the centroid to the centre of the image; the galaxy shape to the true value; the flux to the sum of the pixel intensity values; and the integral defining $T$ is approximated as a discrete sum over pixels. We have verified that this does not bias our results. 
\item Starting from this point, a damped least-squares fit via the Levenberg-Marquardt (LM) algorithm \citep{Levenberg:1944, Marquardt:1963} yields an approximate guess for the likelihood maximum and the covariance of the full distribution; the latter is found via computing the Hessian matrix of the likelihood. 
\item We use this information to draw 50 points from a Gaussian with the given location and shape; these serve as starting values for the independent walkers of a Monte Carlo Markov Chain (MCMC) sampling procedure. {\sc Ngmix} employs the {\tt emcee}\footnote{See \url{dfm.io/emcee/current/}} implementation of affine invariant Monte-Carlo ensemble sampling \citep{ForemanMackey/etal:2012,GoodmanWeare:2010}. For burn-in we sample each walker for 300 steps, and then sample the target distribution for 200 steps. 
\item The ellipticity error as reported by {\sc ngmix}, $\Delta\epsilon$, is given by the standard deviation of the ellipticity $|\epsilon|$ measured from a de-correlated chain of where every tenth point of the original chain is retained.
\end{itemize}

Our choices for the number of walkers, burn-in steps, and sampling steps have been scrutinised and validated: to this end we choose a representative subset of 1024 simulated galaxies, and repeat the three fitting procedures (LSST, Euclid, Joint-Pixel) 32 times, using the same simulated images but different starting conditions for the MCMC walkers. 
We then compute the Gelman-Rubin $R$ statistic for each galaxy \citep{GelmanRubin:1992}. For each of the three fits, the mean, median, and 95th percentile of all 1024 values of $R$ are within one percent of unity. Thus we ensure proper ergodicity and mixing, and avoid spurious early convergence, which would result in an underestimation of the distribution width. 

Both LM and MCMC use the following priors on the sampling parameters: $c_1$ and $c_2$ have Gaussian priors with $\mu$ corresponding to the centre of the image and $\sigma$ equal to half of its angular side length. The joint prior for $\epsilon_1$ and $\epsilon_2$ is a bivariate uniform density supported on the interior of the 2D unit disk. We use wide uniform positive priors on the galaxy size $T$, constraining it to $[0,10^5]$, and the galaxy flux $F$, constraining it to $[0,10^7]$.

For catalogue-level combination, we calculate a weighted average of the LSST and Euclid measurements for each galaxy,
\begin{equation}
\epsilon_{i}^\text{CatComb}=\frac{w^\text{LSST}\epsilon_{i}^\text{LSST}+w^\text{Euclid}\epsilon_{i}^\text{Euclid}}{w^\text{LSST}+w^\text{Euclid}}
\end{equation}
where $i=1,2$ indexes the ellipticity components, and we use the {\sc ngmix} estimated ellipticity errors, $\Delta \epsilon$, to determine inverse-variance weights for each galaxy
\begin{equation}
w^\text{D}= \left(\Delta \epsilon^\text{D}\right)^{-2}\quad\text{for D = LSST, Euclid}.
\label{eqn:weight}
\end{equation}
The error for the catalogue-level combination ellipticity measurement is then initially estimated,  for each galaxy,  as
\begin{equation}
\Delta \epsilon^\text{CatComb}= \frac{1}{\sqrt{w^\text{LSST}+w^\text{Euclid}}}\label{eq:cc_err}\,.
\end{equation}

\subsection{Error Estimates and Calibration}
\label{ssec:shearcal}
We found that the standard deviations of the de-correlated chains produced by {\sc ngmix}, $\Delta\epsilon$, underestimated the true uncertainty in the ellipticity measurements $\delta\epsilon$, particularly in the regime of faint and small galaxies. Error estimates depend on the choice of shape measurement software, and the adopted initialisation strategy, and as they are typically only used to weigh individual galaxies, see for example equation~\ref{eqn:weight}, their measurement need not always be accurate.   In this analysis, however, we wish to use the shape measurement errors in order to quantify the gain in statistical power for different combinations of surveys, and so we require an accurate measurement of the errors.

To assess the accuracy and precision of the four different ellipticity measurements \{LSST, Euclid, CatComb, JointPix\} we compute the calibration corrections and the ellipticity uncertainty $\delta\epsilon$ directly from the image simulations.  We model each component of the observed ellipticity $\epsilon^\text{D}$, where $\text{D}$ denotes the four measurements, as
 \begin{equation}
 \epsilon_i^\text{D}=(1+m^\text{D})\epsilon_i^\text{true}+c^\text{D}+\eta_i^\text{D}\, .
 \label{eq:biascalibration}
 \end{equation}
Here $\eta_i$ is the shape measurement noise error that we model as random variable with expectation value zero and variance $(\delta \epsilon^\text{D})^2$.   
As the simulated PSF is isotropic, $m$ and $c$, the multiplicative and additive calibration corrections respectively, are assumed to be the same for both the $i=1$ and $i=2$ ellipticity components.

To determine the precision of each survey, $\delta \epsilon^\text{D}$, we calculate $m^\text{D}$, $c^\text{D}$ via a fitting procedure that replicates the calibration algorithm in \cite{Miller/etal:2013}. In order to minimise shape noise in our calibration measurements, we bin the fitted galaxies in 30 equal-percentile bins in input magnitude, and 12 equal-width bins in input log-size.   For each of the 360 two-dimensional bins, we perform the following procedure:
\begin{itemize}
\item We further subdivide the galaxies into 20 equal-percentile bins of input shear. Every galaxy is entered into two bins of true shear, since $g_1$ and $g_2$ are treated as independent measures of shear for the purpose of this calibration procedure. 
\item For each shear bin we calculate the average true input shear $\bar{g}^\text{true}$ and the ellipticity-noise-free observed shear $\bar{g}^\text{obs,D}$ as 
\begin{align}
	&\bar{g}^\text{true}=\frac{\sum_j\omega_jg_{j,\alpha}}{\sum_j \omega_j}\label{gtrue}\\
	&\bar{g}^\text{obs,D}=\frac{\sum_j \omega_j(\epsilon^\text{D}_{j,\alpha}-\epsilon^\text{true}_{j,\alpha})}{\sum_j \omega_j}\label{eq:gobs} \,
\end{align}
where the summations are performed over all galaxies $j$ in the shear bin; the index $\alpha=1$ or $2$ depends on whether shear component $g_1$ or $g_2$ has been entered into the bin. The optimal shear calibration weights $\omega_j$ 
\begin{equation}
	\omega_j=\frac{1}{\left(\Delta\epsilon_j^\text{D}\right)^2+\sigma_\text{SN}^2} \, ,
	\label{eqn:optweights}
\end{equation}
combine the estimated uncertainty in the ellipticity measurement from {\sc ngmix}, $\Delta\epsilon_j^\text{D}$, with the intrinsic shape noise per ellipticity component $\sigma_\text{SN} = 0.28$, which we measure directly from the input COSMOS ellipticity catalogue.
\item We calculate an error $\sigma^g$ on the measured shear $\bar{g}^\text{obs,D}$, for each shear bin, using 50 bootstrap realisations.  
\item Collecting all shear bins pertaining to one bin in magnitude and size, we perform a weighted least-squares fit of the model
\begin{equation}
g^\text{obs,D}=(1+m^\text{D})g^\text{true}+c^\text{D}
\end{equation} 
to the data $\left\{\left(\bar g^\text{true},\,\bar g^\text{obs,D},\,\sigma^g\right)\right\}$. This results in fitted values for $m^\text{D}$ and $c^\text{D}$ including their measurement uncertainties for each bin in magnitude and size.
\item The ellipticity measurement uncertainty $\delta\epsilon$ for all galaxies in each magnitude-size-bin is then estimated from the standard deviation of this difference between the true and calibrated shear in the 20 shear bins.
\end{itemize}

To determine the accuracy of each survey, we repeat the calibration procedure on the same fitted ellipticity values $\epsilon^\text{D}$, but now with our robust error bars $\delta\epsilon^\text{D}$ in place of the {\sc ngmix} error bars $\Delta\epsilon^\text{D}$ in equation~\ref{eqn:optweights}. To reduce the noise in the fitted values for $m$ and $c$, we adopt a coarser grid such that each 2D bin contains more galaxies: we split the galaxies into four bins of true magnitude with edges 20 -- 23 -- 24 -- 24.6 -- 25.2; these values are chosen such that they contain similar numbers of galaxies. We further divide into six bins of true size which are logarithmically spaced between 0.05 arcsec and 3 arcsec. For each of the 32 bins in magnitude and size we again subdivide the galaxies into 20 equal-percentile bins of true shear.

\subsection{Effective number density}

When comparing the capability of different galaxy imaging surveys to constrain gravitational shear, a common and useful quantity is the {\it effective number density} of galaxies $n_\text{eff}$ \citep[see][]{Albrecht/etal:2006, Heymans/etal:2012, Chang/etal:2013, Kuijken/etal:2015}. The noise in shear estimation stems from two sources: the individual measurement uncertainty in the shape of each imaged galaxy, and the intrinsic scatter in the ellipticities of the source galaxies. \cite{Chang/etal:2013} define the effective number density of galaxies for a survey as the number density of perfectly measured galaxies (i.e zero measurement uncertainty) which has the same statistical power when constraining gravitational shear.  With this definition 
\begin{equation}
n_\text{eff}=\frac{1}{\Omega}\sum_j\frac{\sigma^2_\text{SN}}{\left(\delta \epsilon_j\right)^2+\sigma^2_\text{SN}},\label{eq:neff_c13}
\end{equation}
where $\Omega$ is the total survey area.  \cite{Heymans/etal:2012} propose an alternative definition for the case where $\sigma^2_\text{SN}$ and its redshift-dependence is unknown.  As $\sigma^2_\text{SN}$ is however known for our simulated sample we adopt equation~\ref{eq:neff_c13}.  It should be noted that $n_\text{eff}$ is a measure of statistical power only, quantifying the potential precision of a shear survey, not its accuracy.

\begin{figure*}
\centering
\begin{minipage}{0.98\textwidth}
 \centering
 \includegraphics[width=\linewidth]{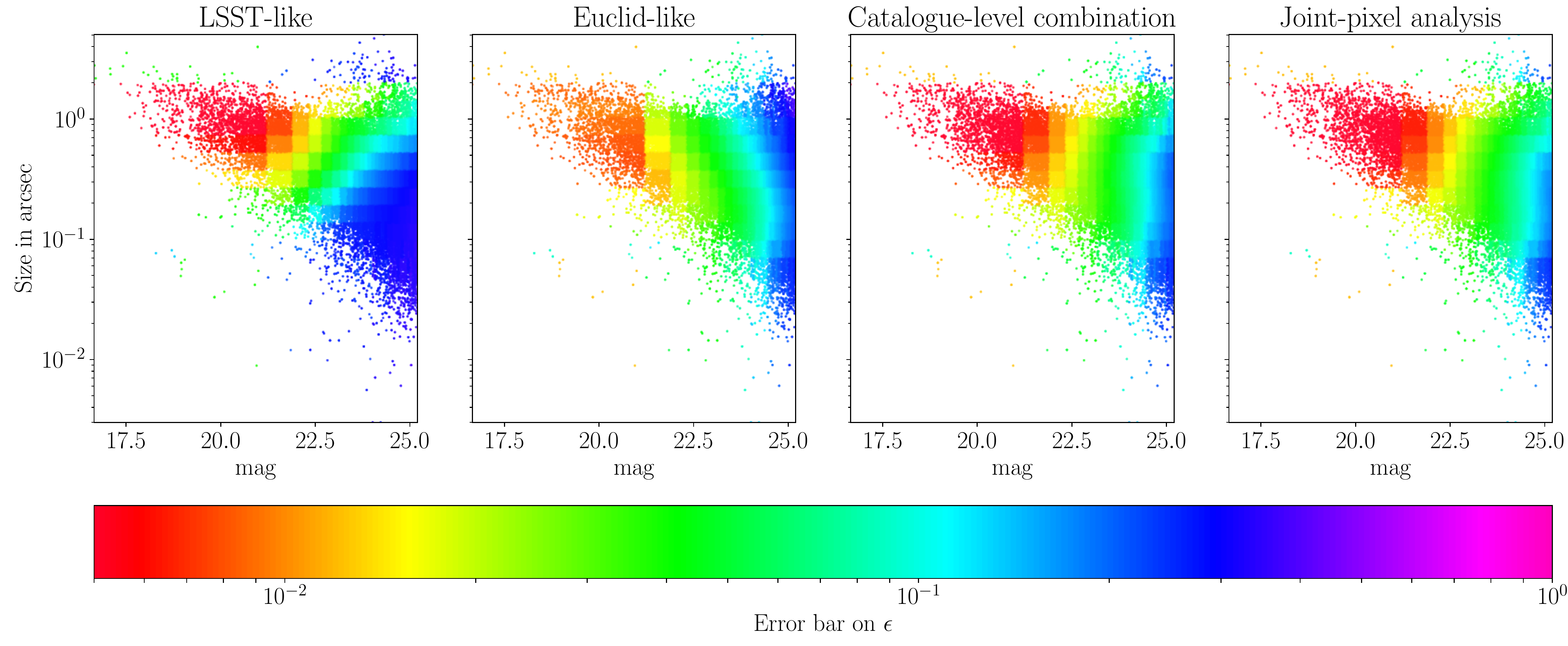}
\end{minipage}
\caption{Error bars on measured ellipticity $\delta \epsilon$, as a function of the input galaxy magnitude and size. {\it From left to right}: LSST-only, Euclid-only, catalogue combination, and joint-pixel analysis.}
\label{fig:e_err}
\end{figure*}
\begin{figure*}
	\centering
	\includegraphics[width=\linewidth]{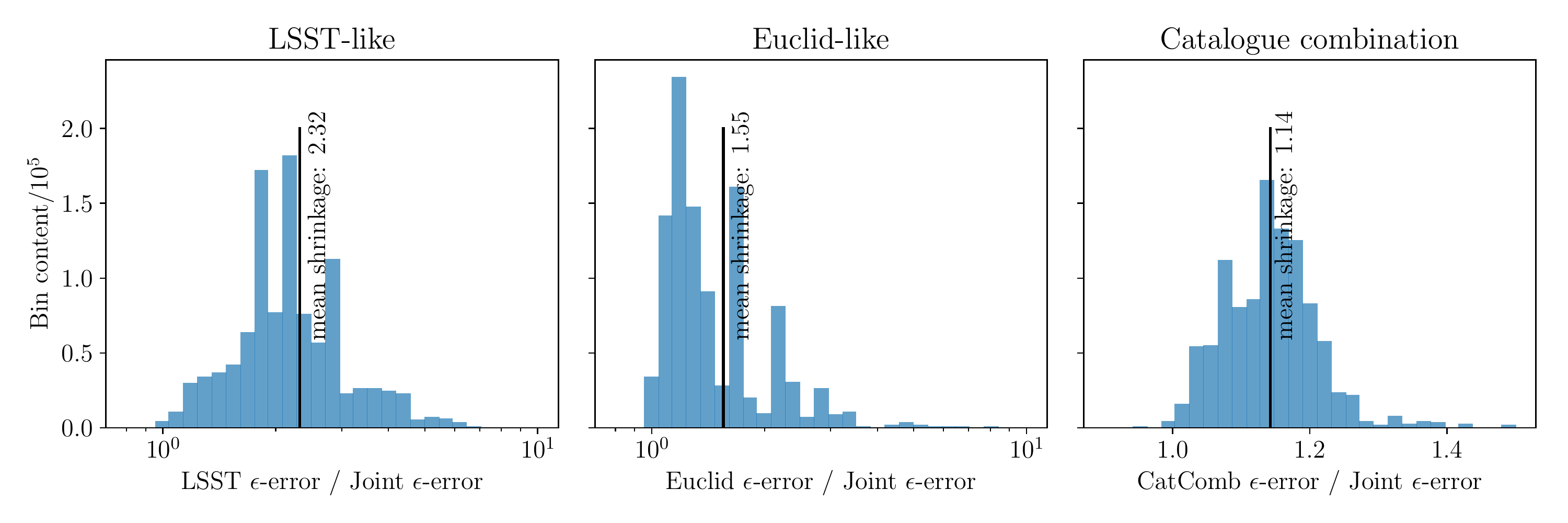}
	\caption{Histograms of shrinkage factors for ellipticity error bars $\delta \epsilon^\text{D}/\delta \epsilon^\text{JointPix}$ when comparing a joint-pixel analysis to the ellipticity measurement $\text{D}$ -- either of LSST, Euclid, and catalogue-level combination ({\it left to right}).}
	\label{fig:improvement_histos}
\end{figure*}

\begin{figure*}
\centering
\begin{minipage}{0.98\textwidth}
 \centering
 \includegraphics[width=\linewidth]{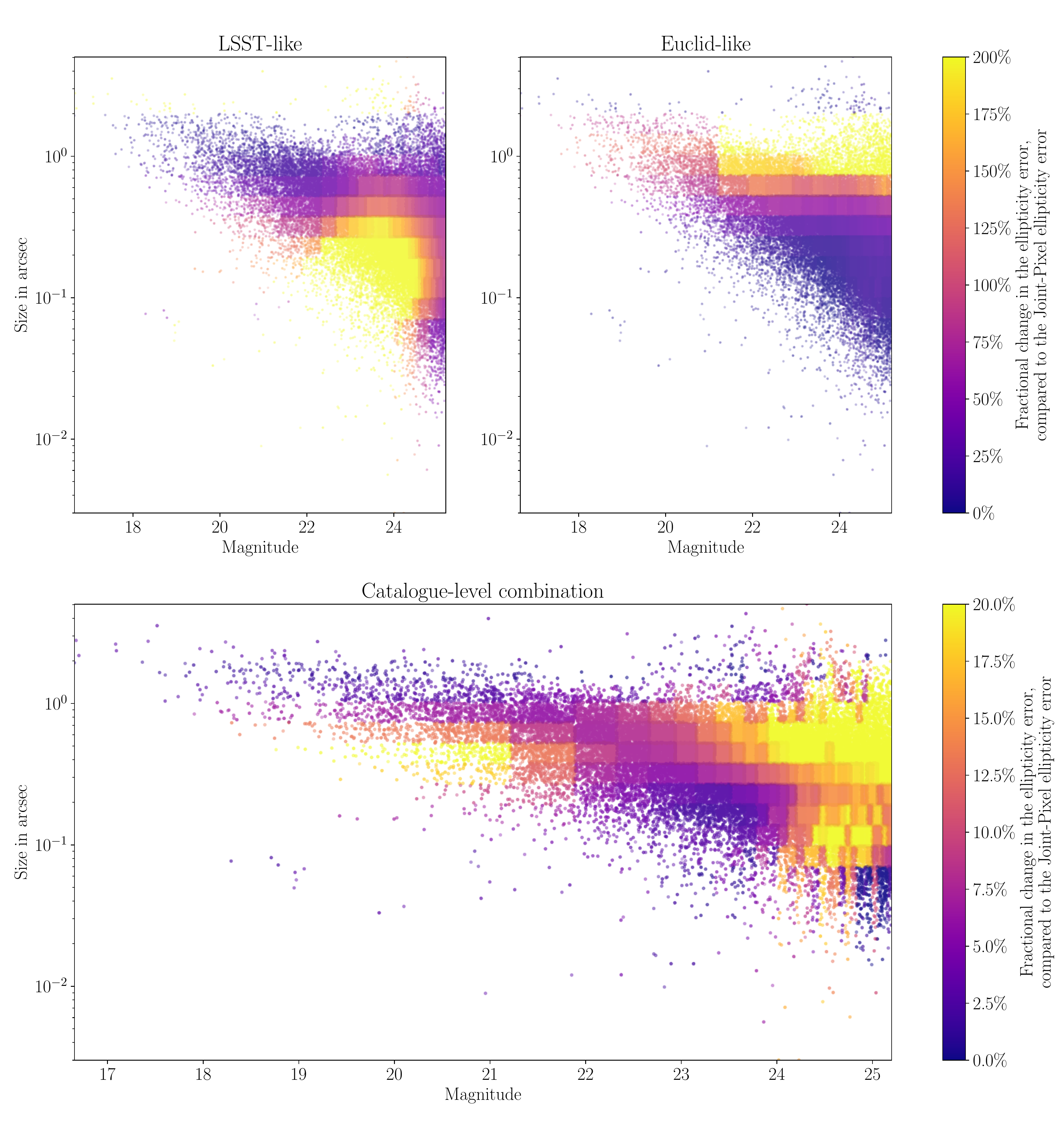}
\end{minipage}
\caption{Precision improvement of a joint-pixel analysis, compared to LSST-only, Euclid-only, and catalogue-level combination. We show the fractional change in the error bar of the total ellipticity, i.e., $(\delta \epsilon^\text{D}-\delta \epsilon^\text{JointPix})/\delta \epsilon^\text{JointPix}$ where $\text{D}\in\{$LSST-like, Euclid-like, catalogue-level combination$\}$.}
\label{fig:e_comp}
\end{figure*}

\begin{figure*}
	\centering
	\includegraphics[width=\linewidth]{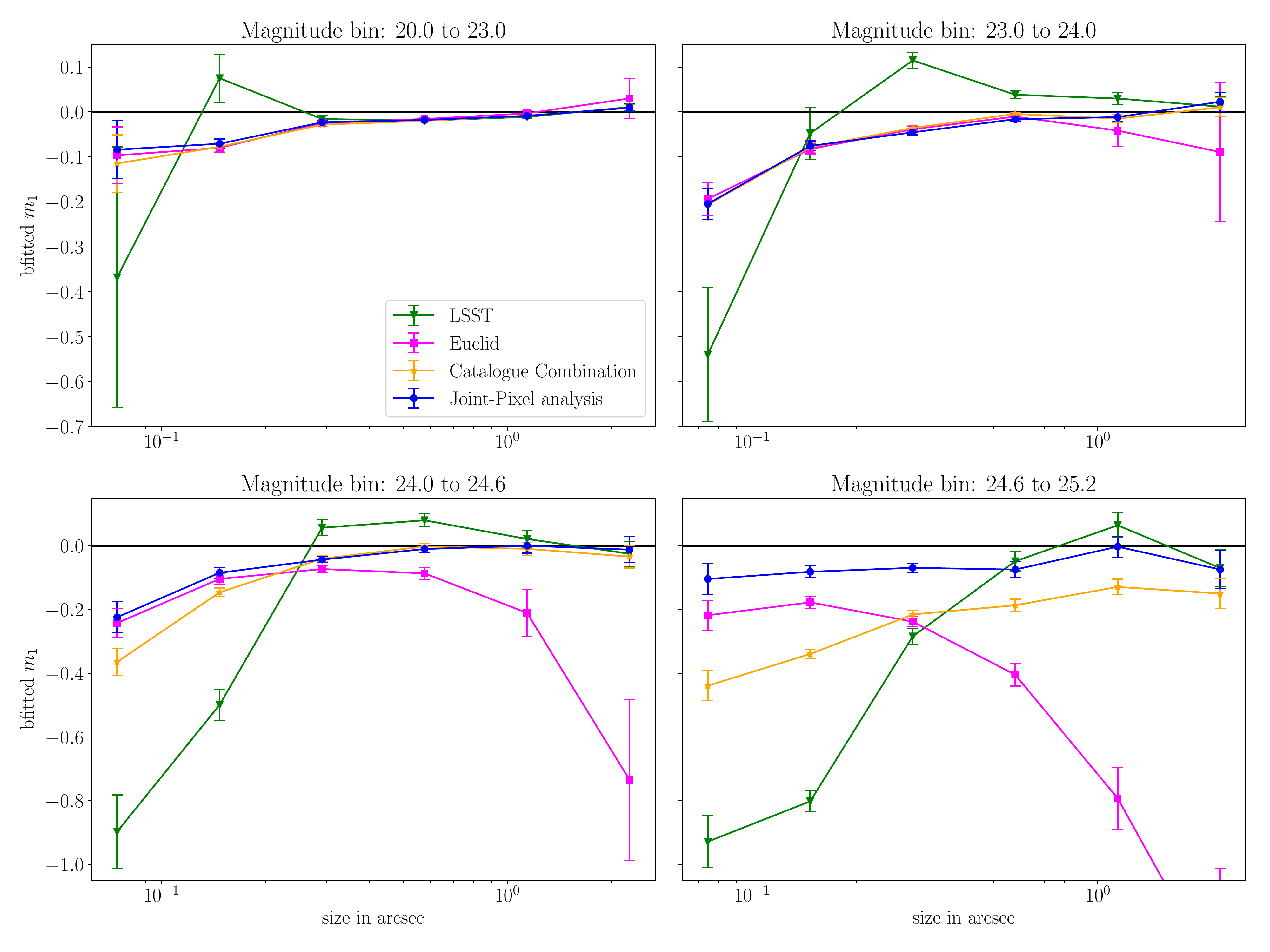}
	\caption{Multiplicative shear calibration factor $m^\text{D}$, where $\text{D}$ is one of $\{\text{LSST-like}$, $\text{Euclid-like}$, $\text{catalogue-level combination}$, $\text{joint-pixel analysis}\}$. We consider four bins in true magnitude and six bins in true galaxy size, and perform the shear calibration procedure described in Section~\ref{ssec:shearcal} to find $m$ and $c$, i.e., the multiplicative and additive calibration parameters. We then plot the fitted $m$ including its 1$\sigma$ uncertainty over the midpoints of the galaxy size bins.}
	\label{fig:shearcal_binned}
\end{figure*}

\section{Results}
\label{sec:results}

We compare precision (Figs.~\ref{fig:e_err}, \ref{fig:improvement_histos}, and \ref{fig:e_comp}) and accuracy (Figure~\ref{fig:shearcal_binned}) of the four different ellipticity measurements \{LSST, Euclid, CatComb, JointPix\} on a set of 1,048,576 simulated galaxy images. Our metrics of improvement are:
\begin{itemize}
 \item the size of the ellipticity error bar $\delta \epsilon^\text{D}$;
 \item the bias of the measured ellipticity $|\epsilon^\text{D}-\epsilon^\text{true}|$;
 \item the effective number of galaxies $N^\text{D}_\text{eff}=n^\text{D}_\text{eff}\Omega$,
\end{itemize}
where $\text{D}$ enumerates the four measurements. 

In Figure~\ref{fig:e_err} we show the ellipticity error bar $\delta\epsilon^\text{D}$ as a function of the true magnitude and size of the input galaxy. The LSST-like and Euclid-like simulations, first and second panels, achieve precise measurements on their own for galaxies that are both bright (magnitude $\lesssim 22$) and large (size $\gtrsim 0.7 \text{ arcsec}$). However, the single-probe measurement error increases outside this region: the LSST-only measurement errors are $\delta\epsilon\lesssim0.05$ out to magnitude $\sim 25$, but only for large galaxies.  The measurement errors degrade rapidly as the galaxy size decreases with $\delta\epsilon\gg0.05$ for small galaxies with size $\lesssim 0.2\text{ arcsec}$. The high-resolution Euclid-like data allows for precise ellipticity measurements for small (size $\sim 0.1 \text{ arcsec}$ and below), but not for faint (magnitude $\gtrsim$ 24) galaxies. This is in agreement with the limits for the LSST resolution and the Euclid $10\sigma$ extended source detection limit, as outlined in \cite{Chang/etal:2013} and \cite{Cropper/etal:2016}.

The third and fourth panels of Figure~\ref{fig:e_err} show the ellipticity error bar of catalogue-level combination and joint-pixel analysis. For both fitting methods, the region of the magnitude-size plane in which ellipticity can be measured with high precision has been extended to include both faint-and-large as well as bright-and-small galaxies. There is, however, a subset of galaxies that are both faint (magnitude $\gtrsim 24.5$) and small (size $\lesssim 0.1 \text{ arcsec}$), whose shape cannot be measured precisely. This is unsurprising, since this would require both high-resolution imaging data and high-depth imaging data.

For a global assessment of the synergy gains, we show in Figure~\ref{fig:improvement_histos} the histogram of the ellipticity error bar shrinkage factors when stepping from either LSST-alone, Euclid-alone, or catalogue combination to joint-pixel fitting. On average, the LSST-only error bars shrink by a factor of 2.32, and the Euclid-only by 1.55, demonstrating the significant precision increase through survey combination. A full joint-pixel analysis exhibits gains over catalogue combination with a shrinkage factor of 1.14 on average in the overall ellipticity measurement errors.  Note that the sampling noise in Figure~\ref{fig:improvement_histos} results from the finite number of 360 galaxy samples, binned by magnitude and size, that are used to make this measurement (see Section~\ref{ssec:shearcal}).

In Figure~\ref{fig:e_comp} we show the fractional precision improvement of a joint-pixel analysis over either of the other three ellipticity measurement methods, in more detail and in its dependency on magnitude and size. The precision enhancements compared to LSST-only can be seen in the regime of galaxies which are small but not too faint (magnitude $\lesssim24$) (due to the addition of the high-resolution Euclid-like data), and for Euclid-only in the regime large and faint galaxies (due to the boost in signal-to-noise from the LSST-like data).

The precision of the catalogue-level combination is equal to the joint-pixel analysis for those galaxies whose shapes can be precisely constrained either by Euclid-only (small and bright) or LSST-only (large and faint). There is, however, a densely populated regime of faint galaxies (magnitude $\gtrsim 24$) of intermediate size -- between the LSST pixel size (0.2 arcsec) and the LSST PSF (0.7 arcsec) -- in which neither single probe can yield a precise shape measurement: it is here that a joint-pixel analysis has its strongest gains over catalogue combination (around 20\%). The same is true for a small population of bright and small galaxies (size between 0.1 and 0.5 arcsec; magnitude $\lesssim 21$). Although each single probe can already yield a precise ellipticity measurement ($\delta\epsilon\lesssim0.01$), the comparison of both combination methods further illustrates: joint-pixel analysis profits over catalogue combination wherever the size of both error bars is comparable.

We assess the accuracy of our measured shapes via the shear calibration procedure described in the last paragraph of Section~\ref{ssec:shearcal}, which uses a coarser grid than was employed for determining the measurement uncertainty $\delta\epsilon$. The results for the shear calibration factor, $m^\text{D}$, are shown in Figure~\ref{fig:shearcal_binned}: for either of the four magnitude bins, the ellipticity measurement from the LSST-like simulation is most strongly biased for galaxies smaller than 0.2 arcsec, whereas the measurements from the Euclid-like simulations exhibit comparably strong biases for low surface-brightness galaxies (faint and large). The majority of the fitted $c$-values are smaller than $5\times10^{-3}$ in absolute value, with a notable exception of the lowest surface-brightness galaxies for the Euclid-only simulation where the zero-centred {\sc ngmix} ellipticity prior dominates the fit resulting in $m = -1.4 \pm 0.4 $, and $c = 0.07 \pm 0.01$. 

Catalogue combination interpolates between LSST-only and Euclid-only, therefore the large LSST bias for small galaxies also increases the bias for the catalogue combination estimate. However, a joint-pixel analysis performs better than either the catalogue combination or the single-telescope measurements, uniting the strengths of both high resolution and high SNR.

\begin{figure}
	\centering
	\includegraphics[width=\linewidth]{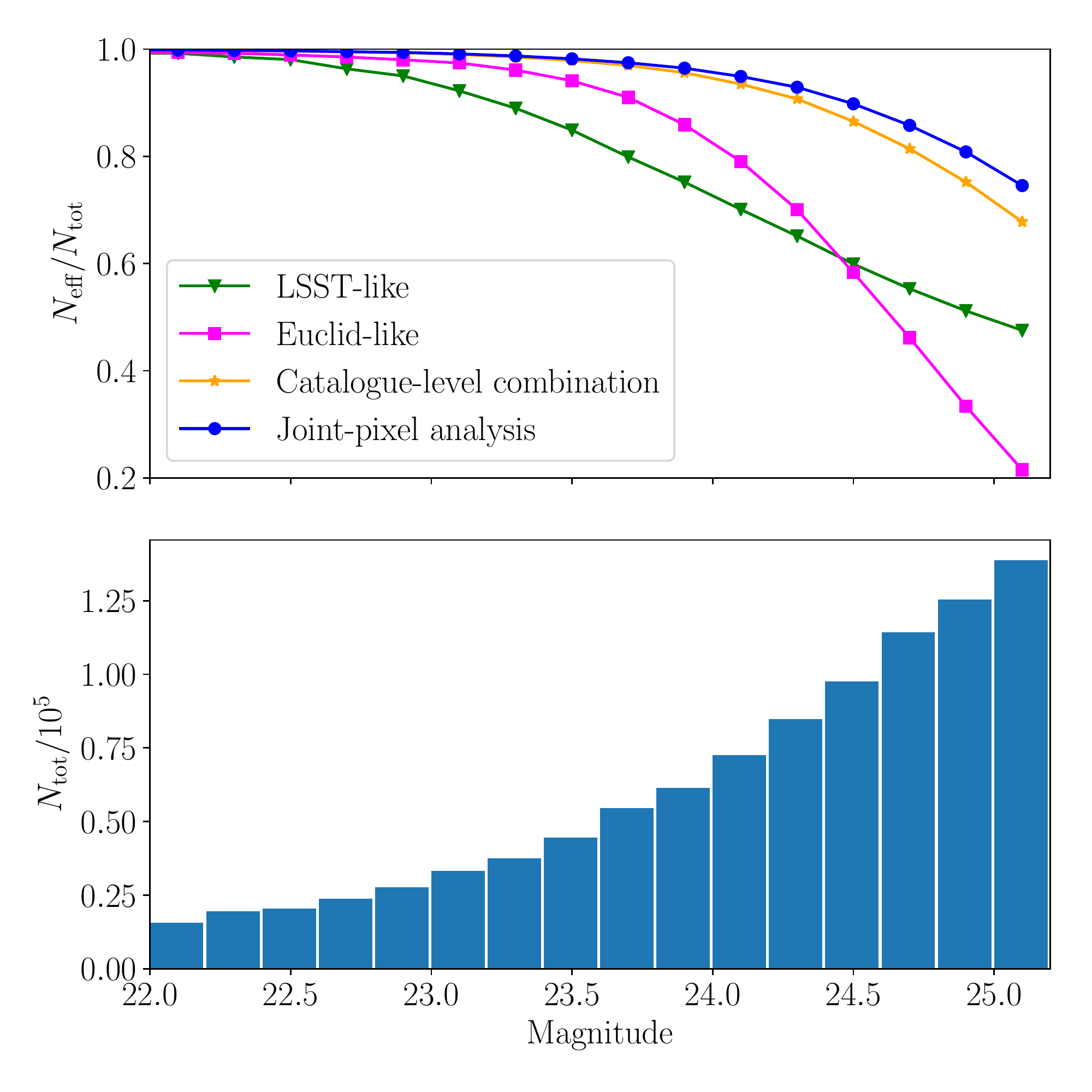}
	\caption{Synergy gains in effective galaxy number for cosmic shear inference. \textit{Top panel}: $N_\text{eff}$ in bins of true magnitude, as a fraction of the total number of galaxies in each bin $N_\text{tot}$. \textit{Lower panel}: total galaxy number per bin $N_\text{tot}$.}
	\label{fig:neff_chang}
\end{figure}

The values for $N_\text{eff}$, as defined by \cite{Chang/etal:2013}, can be computed from our catalogues of error bars $\delta \epsilon^\text{D}$ (see equation~\ref{eq:neff_c13}). To each set of fits, we apply a selection cut at a SNR value of 10. This is to take into account that some galaxies are barely above the detection threshold, and therefore will not enter into a shear measurement analysis. For LSST-like, Euclid-like, and JointPix we use the SNR measured by {\sc ngmix}; the SNR value for catalogue combination is determined via $\text{SNR}_\text{CatComb}=\sqrt{\text{SNR}_\text{LSST}^2+\text{SNR}_\text{Euclid}^2}$. 

We then compute $N_\text{eff}$ for bins of true magnitude; the results are shown in the top panel of Figure~\ref{fig:neff_chang} as fractions of the total number of galaxies in each bin $N_\text{tot}$ before the SNR cut. For the magnitude bins lower than 23, the values for $N_\text{eff}$ are close to $N_\text{tot}$; for the majority of bright simulated galaxies the individual ellipticity measurement error bars are smaller than the intrinsic ellipticity scatter $\left(\sigma_\text{SN}\right)$. Hence the summands in Equation~\ref{eq:neff_c13} are close to unity. For the magnitude bins between 23 and 24.5, the effective galaxy number for the LSST-like simulation does not grow as fast as $N_\text{tot}$ since most of the galaxies are too small to be imaged by LSST. The Euclid-like simulations, which can resolve these galaxies well, outperforms our LSST simulation in this regime. Nevertheless, at the Euclid 10$\sigma$ depth of $i \sim 24.5$ the performance of the Euclid-like simulations falls behind the LSST-like analysis again due to the lack of SNR. At this point, both the joint pixel and catalogue combination analyses are $\sim$45-50$\%$ above either single-instrument survey\footnote{Here, the percentages refer to the gain in the absolute values of $N_\text{eff}$ for the combined probes over single-probe values, not the difference of their fractions to $N_\text{tot}$. It is this percentage that will reflect the potential increase in shear measurement precision.}. In the faint magnitude regime of 24.5-25.2 magnitudes the synergy gain compared to LSST-alone grows to $\sim$45-55$\%$. Since the total number of galaxies is growing steeply due to the larger volume accessible, as shown in the lower panel of Figure~\ref{fig:neff_chang}, the absolute synergy gain in statistical power continues to increase even further.

Both the catalogue combination and the joint pixel-level analysis consistently yield higher numbers for $N_\text{eff}$ and are thus able to constrain gravitational shear more precisely. 

\section{Conclusions and Future Work}
\label{sec:conc}
In this analysis we have demonstrated that a joint analysis of a deep ground-based galaxy imaging survey (like LSST) and a high-resolution space-based survey (like Euclid) will yield significant improvements for both the precision and the accuracy of galaxy shape measurements, compared to the independent analysis of the two surveys. This has the potential to increase the quality of gravitational shear estimation, and thus the statistical constraining power for inference on cosmological parameters.

Survey combination at the joint-pixel level provides a large improvement to shape measurement: the ellipticity error bars shrink by a factor of 2.32 on average compared to LSST-alone (the highest gains coming from small galaxies), and by a factor of 1.55 compared to Euclid-alone (the highest gains coming from faint galaxies). The effective number of galaxies increases by $\sim50\%$ at 25 magnitudes, compounding the statistical precision of both surveys.

A joint pixel analysis has some benefits over combining shear catalogues: it is up to 20\% more precise where neither catalogue provides a good ellipticity measurement; it is also more accurate, facilitating the calibration of measured shapes.  As the statistical power of cosmic shear is dominated by intrinsic ellipticity noise, $\sigma_{\rm SN}$, however, this improvement in shape measurement noise only leads to a $\sim 5$ percent improvement in the effective number density of galaxies for lensing studies at faint magnitudes.

There are a number of caveats to our analysis.  
In order to get a first impression of the improvements to galaxy shape measurement when combining space and ground data, we chose not to use the advanced machinery developed by either the LSST or Euclid collaborations for simulating and analysing realistic images.  We also simplified the challenge of image combination by assuming perfect registration between the two surveys, perfectly known round Gaussian PSFs, and, most importantly, isolated galaxies.
Future work should make use of the powerful tools within the Euclid and LSST collaborations in order to increase the realism of the image simulations in terms of the noise model, morphology, masking, image distortion, and the PSF model. 
Our choice of shape fitting software ({\sc ngmix}) will also impact our conclusions, to some degree, and this analysis should therefore be repeated for a range of different shear measurement and calibration techniques, for example {\sc Metacalibration}, {\sc Im3Shape} and \emph{lens}fit \citep{Huff/Mandelbaum:2017, Zuntz/etal:2013, Miller/etal:2013}.

Our use of isolated galaxies means that we have not investigated what is arguably most important advantage of combining space-based and ground-based data: \emph{deblending}, the splitting of light from close (\emph{blended}) objects into separate sources.  High resolution space-based data will improve the ability of ground-based telescopes to unambiguously discriminate and model close pairs of galaxies in a joint-pixel analysis \citep[see for example][for a review]{Mandelbaum:2017}.  Combining with high-resolution data will also improve star-galaxy separation for ground-based analyses, by distinguishing smaller galaxies from stars more cleanly, leading to an improvement in PSF modelling.   Since we have omitted these major advantages of combined data analysis, we expect that our analysis quantifies only the minimal level of improvement that should be expected when adopting a joint space-ground-pixel shape measurement analysis with the actual improvements likely to be even greater than those presented here.

\section{Acknowledgements}
We thank Bryan R. Gillis, Benjamin Joachimi, David Kirkby and Mark Cropper for helpful discussions, and Erin Sheldon and the GREAT3 team for making their shape measurement and image simulation software publicly available.  We acknowledge support from STFC for UK participation in LSST through grant numbers ST/N002539/1 (RLS) and ST/N002512/1 (CH, JZ).  CH also acknowledges support from the European Research Council under grant number 647112.  This work was carried out in part at the Aspen Center for Physics, which is supported by the National Science Foundation grant PHY-1607611, where CH and JZ were also supported by a grant from the Simons Foundation.



\bibliographystyle{mnras}
\bibliography{LSSTxEuclid} 




\appendix
\section{LSST sky brightness}
In this appendix we describe how we calculate the $i$-band sky background of $s= 975.7 e^-$ per pixel for a single LSST exposure listed in Table~\ref{tab:galsimpars}, following \citealt{Jones:2017}; David Kirkby, (private communication). The flux (in photo-electrons) for a single-exposure duration $T$ (in sec) is given by: 
\begin{equation}
  s=s_0 \, T \times 10^{-0.4(m_\text{sky}-m_0)} \, .
\end{equation} 
The zero point $s_0$, defined as the flux of a magnitude $m_0$ source, is given by $s_0 = \zeta A$ where A is the collecting area and $\zeta$ is a constant which depends on the bandpass.   For the LSST $i$-band we use $m_0 = 24$, $\zeta =0.999 e^-$/(m$^2$ sec), a sky brightness of $m_\text{sky}=20.5$, $T=30$ sec, and an effective telescope diameter of $D = 6.423\mathrm{m}$, which yields $A=32.4 \mathrm{m}^2$.   


 \bsp
\label{lastpage}
\end{document}

%% file: abstract_ch.tex
We demonstrate that a joint analysis of LSST-like ground-based imaging with Euclid-like space-based imaging leads to increased precision and accuracy in galaxy shape measurements.  At galaxy magnitudes of $i \sim 24.5$, a combined survey analysis increases the effective galaxy number density for cosmic shear studies by $\sim 50$ percent in comparison to an analysis of each survey alone.  Using a realistic distribution of galaxy sizes, ellipticities and magnitudes down to $i = 25.2$, we simulate LSST-like and Euclid-like images of over one million isolated galaxies.   We compare the precision and accuracy of the recovered galaxy ellipticities for four different analyses:  LSST-only, Euclid-only, a simultaneous joint-pixel analysis of the two surveys, and a simple catalogue-level survey combination.   In the faint and small-galaxy regime, where neither survey excels alone, we find a $\sim 20$ percent increase in the precision of galaxy shape measurement when we adopt a joint-pixel analysis, compared to a catalogue-level combination.  As the statistical power of cosmic shear is dominated by intrinsic ellipticity noise, however, this improvement in shape measurement noise only leads to a $\sim 5$ percent improvement in the effective number density of galaxies for lensing studies. We view this as the minimum improvement that should be expected from a joint-pixel analysis over a less accurate catalogue-level combination, as the former will also improve the capability of LSST to de-blend close objects.

%% file: introduction.tex
Weak gravitational lensing is a technique which exploits the fact that large structures of invisible dark matter gravitationally deflect light rays, coherently distorting the images of the distant galaxies that we observe behind them.  This lensing effect is directly sensitive to the distribution of matter in the Universe, giving us the rare ability to produce direct comparisons between observations and theories of dark matter and dark energy.   In the absence of systematic errors, weak lensing is recognised as the single most constraining probe of dark energy \citep[see for example][]{Albrecht/etal:2006} and is a primary science driver for two of the major imaging surveys of the 2020's; Euclid\footnote{Euclid: \url{http://www.euclid-ec.org}} and the Large Synoptic Survey Telescope (LSST\footnote{LSST: \url{http://www.lsst.org}}).    

Euclid is a 1.2m space-based telescope that will image 15,000 square degrees in one broad optical band to a depth of 24.5 AB magnitudes ($10\sigma$ extended source), in addition to three near-infrared bands \citep{Laureijs/etal:2011,Cropper/etal:2016}. LSST is a ground-based telescope with an effective mirror diameter of 6.7m. Over ten years it will image $\sim 18,000$ square degrees in six optical bands to a depth of $r \sim 27.5$ \citep[$5\sigma$ point source,][]{Ivezic/etal:2008,Chang/etal:2013}.   The two surveys will commence observations on similar timescales; LSST will collect extremely high signal-to-noise imaging, but will lack resolution, with an average seeing of 0.7 arcsec.  In contrast, Euclid's space-based imaging will have exquisite resolution, but will lack both signal-to-noise at faint magnitudes and the multi-colour optical imaging that LSST provides.   Optimal weak lensing measurements benefit from both high resolution and high signal-to-noise imaging \citep[see for example][]{Massey/etal:2013}, which naturally leads to the suggestion that it is the combination of these two surveys that will provide the optimal measurement of weak lensing for the majority of the galaxy population that are both small and faint (see Figure~\ref{fig:histo}).
In \citet{rhodes} and  \citet{Jain/etal:2015}, the synergy, benefits and challenges associated with analysing Euclid and LSST in concert were reviewed, with the conclusion that collaboration between the surveys will maximise the overall scientific return.  In a first quantitative example, \citet{rhodes} determine the expected decrease in photometric redshift error when Euclid and LSST photometric data is used in tandem.   With improved redshift estimates, the signal-to-noise of the combined weak lensing signal from each survey is shown to increase by $\sim 30$ percent, assuming an overlap between the two surveys of 7000 square degrees.  

In this paper we present a second quantitative example of the benefits of the joint-analysis of Euclid and LSST, focusing on weak lensing shape measurement precision.   We seek to address the question of whether the benefits of a joint-pixel-level shear analysis are sufficiently high to warrant the additional technical complexity that such an analysis would incur, in contrast to adopting a simple catalogue-level combination of shear measurements.   In this first test case we choose to limit our analysis to single-band imaging of isolated galaxies, deferring for future work the challenges of object blends and colour gradients that joint pixel-level measurements would also help to resolve. 

In Section~\ref{sec:methods} we discuss our adopted shape measurement software {\sc ngmix}; we also describe our suite of image simulations for two test cases; an almost noise-free but low-resolution LSST-like survey, and an almost point-spread function (PSF)-free but noisy Euclid-like survey. We present the results of our single-survey, combined catalogue and joint-pixel analysis in Section~\ref{sec:results} and conclude in Section~\ref{sec:conc}.

%% file: LSSTxEuclid.bbl
\begin{thebibliography}{}
\makeatletter
\relax
\def\mn@urlcharsother{\let\do\@makeother \do\$\do\&\do\#\do\^\do\_\do\%\do\~}
\def\mn@doi{\begingroup\mn@urlcharsother \@ifnextchar [ {\mn@doi@}
  {\mn@doi@[]}}
\def\mn@doi@[#1]#2{\def\@tempa{#1}\ifx\@tempa\@empty \href
  {http://dx.doi.org/#2} {doi:#2}\else \href {http://dx.doi.org/#2} {#1}\fi
  \endgroup}
\def\mn@eprint#1#2{\mn@eprint@#1:#2::\@nil}
\def\mn@eprint@arXiv#1{\href {http://arxiv.org/abs/#1} {{\tt arXiv:#1}}}
\def\mn@eprint@dblp#1{\href {http://dblp.uni-trier.de/rec/bibtex/#1.xml}
  {dblp:#1}}
\def\mn@eprint@#1:#2:#3:#4\@nil{\def\@tempa {#1}\def\@tempb {#2}\def\@tempc
  {#3}\ifx \@tempc \@empty \let \@tempc \@tempb \let \@tempb \@tempa \fi \ifx
  \@tempb \@empty \def\@tempb {arXiv}\fi \@ifundefined
  {mn@eprint@\@tempb}{\@tempb:\@tempc}{\expandafter \expandafter \csname
  mn@eprint@\@tempb\endcsname \expandafter{\@tempc}}}

\bibitem[\protect\citeauthoryear{{Albrecht} et~al.,}{{Albrecht}
  et~al.}{2006}]{Albrecht/etal:2006}
{Albrecht} A.,  et~al., 2006, ArXiv Astrophysics e-prints, \href
  {http://adsabs.harvard.edu/abs/2006astro.ph..9591A} {}

\bibitem[\protect\citeauthoryear{{Chang} et~al.,}{{Chang}
  et~al.}{2013}]{Chang/etal:2013}
{Chang} C.,  et~al., 2013, \mn@doi [\mnras] {10.1093/mnras/stt1156}, \href
  {http://adsabs.harvard.edu/abs/2013MNRAS.434.2121C} {434, 2121}

\bibitem[\protect\citeauthoryear{{Cropper} et~al.,}{{Cropper}
  et~al.}{2016}]{Cropper/etal:2016}
{Cropper} M.,  et~al., 2016, in Space Telescopes and Instrumentation 2016:
  Optical, Infrared, and Millimeter Wave. p. 99040Q (\mn@eprint {arXiv}
  {1608.08603}), \mn@doi{10.1117/12.2234739}

\bibitem[\protect\citeauthoryear{{Foreman-Mackey}, {Hogg}, {Lang}  \&
  {Goodman}}{{Foreman-Mackey} et~al.}{2013}]{ForemanMackey/etal:2012}
{Foreman-Mackey} D.,  {Hogg} D.~W.,  {Lang} D.,   {Goodman} J.,  2013, \mn@doi
  [\pasp] {10.1086/670067}, \href
  {http://adsabs.harvard.edu/abs/2013PASP..125..306F} {125, 306}

\bibitem[\protect\citeauthoryear{{Gelman} \& {Rubin}}{{Gelman} \&
  {Rubin}}{1992}]{GelmanRubin:1992}
{Gelman} A.,  {Rubin} D.~B.,  1992, \mn@doi [Statistical Science]
  {10.1214/ss/1177011136}, \href
  {http://adsabs.harvard.edu/abs/1992StaSc...7..457G} {7, 457}

\bibitem[\protect\citeauthoryear{{Goodman} \& {Weare}}{{Goodman} \&
  {Weare}}{2010}]{GoodmanWeare:2010}
{Goodman} J.,  {Weare} J.,  2010, \mn@doi [Communications in Applied
  Mathematics and Computational Science, Vol.~5, No.~1, p.~65-80, 2010]
  {10.2140/camcos.2010.5.65}, \href
  {http://adsabs.harvard.edu/abs/2010CAMCS...5...65G} {5, 65}

\bibitem[\protect\citeauthoryear{{Heymans} et~al.,}{{Heymans}
  et~al.}{2012}]{Heymans/etal:2012}
{Heymans} C.,  et~al., 2012, \mn@doi [\mnras]
  {10.1111/j.1365-2966.2012.21952.x}, \href
  {http://adsabs.harvard.edu/abs/2012MNRAS.427..146H} {427, 146}

\bibitem[\protect\citeauthoryear{{Hogg} \& {Lang}}{{Hogg} \&
  {Lang}}{2013}]{Hogg/Lang:2012}
{Hogg} D.~W.,  {Lang} D.,  2013, \mn@doi [\pasp] {10.1086/671228}, \href
  {http://adsabs.harvard.edu/abs/2013PASP..125..719H} {125, 719}

\bibitem[\protect\citeauthoryear{{Huff} \& {Mandelbaum}}{{Huff} \&
  {Mandelbaum}}{2017}]{Huff/Mandelbaum:2017}
{Huff} E.,  {Mandelbaum} R.,  2017, preprint, \href
  {http://adsabs.harvard.edu/abs/2017arXiv170202600H} {} (\mn@eprint {arXiv}
  {1702.02600})

\bibitem[\protect\citeauthoryear{{Ivezic} et~al.,}{{Ivezic}
  et~al.}{2008}]{Ivezic/etal:2008}
{Ivezic} Z.,  et~al., 2008, preprint, \href
  {http://adsabs.harvard.edu/abs/2008arXiv0805.2366I} {} (\mn@eprint {arXiv}
  {0805.2366})

\bibitem[\protect\citeauthoryear{{Jain} et~al.,}{{Jain}
  et~al.}{2015}]{Jain/etal:2015}
{Jain} B.,  et~al., 2015, preprint, \href
  {http://adsabs.harvard.edu/abs/2015arXiv150107897J} {} (\mn@eprint {arXiv}
  {1501.07897})

\bibitem[\protect\citeauthoryear{Jarvis et~al.,}{Jarvis
  et~al.}{2016}]{Jarvis/etal:2016}
Jarvis M.,  et~al., 2016, \mn@doi [Monthly Notices of the Royal Astronomical
  Society] {10.1093/mnras/stw990}, 460, 2245

\bibitem[\protect\citeauthoryear{{Jones}}{{Jones}}{2017}]{Jones:2017}
{Jones} L.,  2017, SM Technical Note, SMTN-002

\bibitem[\protect\citeauthoryear{{Kuijken} et~al.,}{{Kuijken}
  et~al.}{2015}]{Kuijken/etal:2015}
{Kuijken} K.,  et~al., 2015, \mn@doi [\mnras] {10.1093/mnras/stv2140}, \href
  {http://adsabs.harvard.edu/abs/2015MNRAS.454.3500K} {454, 3500}

\bibitem[\protect\citeauthoryear{{LSST Science Collaboration} et~al.,}{{LSST
  Science Collaboration} et~al.}{2009}]{LSSTScienceBook/2009}
{LSST Science Collaboration} et~al., 2009, preprint, \href
  {http://adsabs.harvard.edu/abs/2009arXiv0912.0201L} {} (\mn@eprint {arXiv}
  {0912.0201})

\bibitem[\protect\citeauthoryear{{Lackner} \& {Gunn}}{{Lackner} \&
  {Gunn}}{2012}]{Lackner/Gunn:2012}
{Lackner} C.~N.,  {Gunn} J.~E.,  2012, \mn@doi [\mnras]
  {10.1111/j.1365-2966.2012.20450.x}, \href
  {http://adsabs.harvard.edu/abs/2012MNRAS.421.2277L} {421, 2277}

\bibitem[\protect\citeauthoryear{{Laureijs} et~al.,}{{Laureijs}
  et~al.}{2011}]{Laureijs/etal:2011}
{Laureijs} R.,  et~al., 2011, preprint, \href
  {http://adsabs.harvard.edu/abs/2011arXiv1110.3193L} {} (\mn@eprint {arXiv}
  {1110.3193})

\bibitem[\protect\citeauthoryear{{Leauthaud} et~al.,}{{Leauthaud}
  et~al.}{2007}]{Leauthaud/etal:2007}
{Leauthaud} A.,  et~al., 2007, \mn@doi [\apjs] {10.1086/516598}, \href
  {http://adsabs.harvard.edu/abs/2007ApJS..172..219L} {172, 219}

\bibitem[\protect\citeauthoryear{Levenberg}{Levenberg}{1944}]{Levenberg:1944}
Levenberg K.,  1944, Quarterly Journal of Applied Mathmatics, II, 164

\bibitem[\protect\citeauthoryear{{Mandelbaum}}{{Mandelbaum}}{2018}]{Mandelbaum:2017}
{Mandelbaum} R.,  2018, \mn@doi [Annual Review of Astronomy and Astrophysics]
  {10.1146/annurev-astro-081817-051928}, \href
  {https://ui.adsabs.harvard.edu/\#abs/2018ARA&A..56..393M} {56, 393}

\bibitem[\protect\citeauthoryear{{Mandelbaum} et~al.,}{{Mandelbaum}
  et~al.}{2014}]{Mandelbaum/etal:2014}
{Mandelbaum} R.,  et~al., 2014, \mn@doi [\apjs] {10.1088/0067-0049/212/1/5},
  \href {http://adsabs.harvard.edu/abs/2014ApJS..212....5M} {212, 5}

\bibitem[\protect\citeauthoryear{Marquardt}{Marquardt}{1963}]{Marquardt:1963}
Marquardt D.~W.,  1963, \mn@doi [Journal of the Society for Industrial and
  Applied Mathematics] {10.1137/0111030}, 11, 431

\bibitem[\protect\citeauthoryear{{Massey} et~al.,}{{Massey}
  et~al.}{2013}]{Massey/etal:2013}
{Massey} R.,  et~al., 2013, \mn@doi [\mnras] {10.1093/mnras/sts371}, \href
  {http://adsabs.harvard.edu/abs/2013MNRAS.429..661M} {429, 661}

\bibitem[\protect\citeauthoryear{{Meyers} \& {Burchat}}{{Meyers} \&
  {Burchat}}{2015}]{Meyers:2014}
{Meyers} J.~E.,  {Burchat} P.~R.,  2015, \mn@doi [\apj]
  {10.1088/0004-637X/807/2/182}, \href
  {http://adsabs.harvard.edu/abs/2015ApJ...807..182M} {807, 182}

\bibitem[\protect\citeauthoryear{{Miller} et~al.,}{{Miller}
  et~al.}{2013}]{Miller/etal:2013}
{Miller} L.,  et~al., 2013, \mn@doi [\mnras] {10.1093/mnras/sts454}, \href
  {http://adsabs.harvard.edu/abs/2013MNRAS.429.2858M} {429, 2858}

\bibitem[\protect\citeauthoryear{{Rhodes} et~al.,}{{Rhodes}
  et~al.}{2017}]{rhodes}
{Rhodes} J.,  et~al., 2017, \mn@doi [\apjs] {10.3847/1538-4365/aa96b0}, \href
  {http://adsabs.harvard.edu/abs/2017ApJS..233...21R} {233, 21}

\bibitem[\protect\citeauthoryear{{Rowe} et~al.,}{{Rowe}
  et~al.}{2015}]{Rowe/etal:2015}
{Rowe} B.~T.~P.,  et~al., 2015, \mn@doi [Astronomy and Computing]
  {10.1016/j.ascom.2015.02.002}, \href
  {http://adsabs.harvard.edu/abs/2015A%26C....10..121R} {10, 121}

\bibitem[\protect\citeauthoryear{{Seitz} \& {Schneider}}{{Seitz} \&
  {Schneider}}{1997}]{SeitzSchneider:1997}
{Seitz} C.,  {Schneider} P.,  1997, \aap, \href
  {http://adsabs.harvard.edu/abs/1997A%26A...318..687S} {318, 687}

\bibitem[\protect\citeauthoryear{{S{\'e}rsic}}{{S{\'e}rsic}}{1963}]{sersic:1963}
{S{\'e}rsic} J.~L.,  1963, Boletin de la Asociacion Argentina de Astronomia La
  Plata Argentina, \href {http://adsabs.harvard.edu/abs/1963BAAA....6...41S}
  {6, 41}

\bibitem[\protect\citeauthoryear{{Sheldon}}{{Sheldon}}{2014}]{Sheldon:2014}
{Sheldon} E.~S.,  2014, \mn@doi [\mnras] {10.1093/mnrasl/slu104}, \href
  {http://adsabs.harvard.edu/abs/2014MNRAS.444L..25S} {444, L25}

\bibitem[\protect\citeauthoryear{{Zuntz}, {Kacprzak}, {Voigt}, {Hirsch}, {Rowe}
   \& {Bridle}}{{Zuntz} et~al.}{2013}]{Zuntz/etal:2013}
{Zuntz} J.,  {Kacprzak} T.,  {Voigt} L.,  {Hirsch} M.,  {Rowe} B.,   {Bridle}
  S.,  2013, \mn@doi [\mnras] {10.1093/mnras/stt1125}, \href
  {http://adsabs.harvard.edu/abs/2013MNRAS.434.1604Z} {434, 1604}

\bibitem[\protect\citeauthoryear{{Zuntz} et~al.,}{{Zuntz}
  et~al.}{2018}]{Zuntz/etal:2017}
{Zuntz} J.,  et~al., 2018, \mn@doi [\mnras] {10.1093/mnras/sty2219}, \href
  {https://ui.adsabs.harvard.edu/\#abs/2018MNRAS.481.1149Z} {481, 1149}

\makeatother
\end{thebibliography}
